\documentclass[%
 reprint,
superscriptaddress,
nobibnotes,
 amsmath,amssymb,
 aps,
 prl,
]{revtex4-2}

\usepackage{graphicx}
\usepackage{dcolumn}
\usepackage{bm}
\usepackage{bbm}
\usepackage[colorlinks=true, allcolors=blue]{hyperref}
\usepackage{braket}
\usepackage{mathtools}
\usepackage{stmaryrd}
\usepackage[ruled,linesnumbered,vlined,titlenotnumbered]{algorithm2e} 
\usepackage{tikz}
\usepackage{tikz-cd}
\usepackage{amsthm}
\usepackage[normalem]{ulem}
\usepackage[dvipsnames]{xcolor}
\usepackage{comment}
\usepackage{adjustbox}
\usepackage[caption=false]{subfig}
\theoremstyle{definition}

\theoremstyle{remark}

\renewcommand{\bar}{\overline}

\renewcommand{\le}{\leqslant}
\renewcommand{\ge}{\geqslant}

\newcommand{\Tr}{\operatorname{Tr}}

\newcommand{\dkap}{\delta\kern-1.25pt\varkappa}

\newcommand{\ZZ}{\mathbb{Z}}

\newcommand{\calP}{\mathcal{P}}

\makeatletter
\newcommand{\RemoveAlgoNumber}{\renewcommand{\fnum@algocf}{\AlCapSty{\AlCapFnt\algorithmcfname}}}
\newcommand{\RevertAlgoNumber}{\algocf@resetfnum}
\makeatother

\begin{document}

\title{Spinal reflected entropy reveals logical magic}

\author{Zong-Yue Hou}
\affiliation{Institute for Interdisciplinary Information Sciences, Tsinghua University, Beijing, 100084, China}

\author{Yuzhen Zhang}
\affiliation{Department of Physics, University of California, Santa Barbara, California, 93106, USA}

\author{Zhi-Cheng Yang}
\affiliation{School of Physics, Peking University, Beijing 100871, China}
\affiliation{Center for High Energy Physics, Peking University, Beijing 100871, China}

\author{Yingfei Gu}
\email[E-mail: ]{guyingfei@tsinghua.edu.cn}
\affiliation{Chen-Ning Yang Institute for Advanced Study, Tsinghua University, Beijing, 100084, China}
\affiliation{Beijing Key Laboratory of Cold Atom Quantum Computation, Tsinghua University, Beijing, 100084, China}

\date{\today}

\begin{abstract}
Entanglement and magic are usually regarded as independent quantum resources. Here we
connect the two in the toric code, where the stabilizer R\'enyi entropy of the stored
logical state, a widely used measure of magic, acquires a direct interpretation in terms
of entanglement. We introduce the \emph{spinal reflected entropy}, defined on a
neighborhood of a spine of the underlying surface as the reflected entropy between a disk
around the vertex of the spine and bands along its noncontractible loops. It obeys an
area law with a topological correction and contains one additional term that carries all
the dependence on the stored state. On a torus, this term equals the stabilizer R\'enyi
entropy of the stored state plus the logarithm of the ground-state degeneracy. The
mechanism is topological. On the surface sewn to its reflected copy, the bands close into
tubes, logical Pauli operators become anyon charges threading them, and these charges are
distributed as the outcomes of Bell sampling of the stored state. The result extends to
surfaces of higher genus and to general two-dimensional topological stabilizer codes on
qubits. Logical magic thus becomes a geometrically defined entanglement quantity, one that
shallow circuits preserving the code space cannot change.
\end{abstract}

\maketitle
\raggedbottom

\emph{Introduction.---}
Entanglement and magic are usually regarded as independent resources, because
they are defined with respect to different structures. Entanglement refers to
a division of the system into parts and quantifies the correlations between
them. Magic, or nonstabilizerness, refers instead to the Pauli group and
quantifies the departure from stabilizer states. Stabilizer states acted on
by Clifford gates can be simulated efficiently on a classical
computer~\cite{Gottesman:1998hu}, and magic is the additional resource that
makes Clifford operations universal~\cite{Bravyi:2004isx}. Neither
resource implies the other. Single-qubit gates leave entanglement unchanged
but can create magic, whereas Clifford gates leave magic unchanged but can
create entanglement. The interplay of the two continues to attract
considerable research
interest~\cite{hou2026stabilizer, iannotti2025entanglement, jplh-zl35, PhysRevB.110.045101}.

The contrast is also present at the level of logical qubits, protected by a
quantum error-correcting code such as the toric
code~\cite{kitaev2003fault}. On a surface with handles, the toric code has a
degenerate ground space, which serves as the code space. Its logical
operators are string operators around the handles, so local noise can neither
read nor alter the stored state, and the system is a robust quantum
memory~\cite{Dennis:2001nw}.
In this setting, the two resources appear to sit at different levels. The
standard entanglement signature belongs to the code. For a disk-shaped
region with boundary length $L$, the entanglement entropy of a ground state
has the form $S=\alpha L-\gamma$, where the topological entanglement entropy
$\gamma=\log\mathcal D$ is fixed by the total quantum dimension $\mathcal D$
of the anyon theory~\cite{kitaev2006topological,levin2006detecting}.
A disk sees the same
entanglement in every ground state, so the topological entanglement entropy
is blind to which logical state is stored. Magic, by contrast, belongs to the
stored state. The relevant Pauli operators are the logical ones, the string
operators around the handles, and magic depends on the expectation values of
all of them at once.

Entanglement is not entirely blind to the stored state. When the cut wraps a
handle, the entropy depends on the stored state through its weights on
sectors of definite anyon flux across the
cut~\cite{dong2008topological,zhang2012quasiparticle,bonderson2017anyonic}.
But a single cut sees these weights only in the basis selected by its
geometry, the eigenbasis of the string operators parallel to it. Cuts along
different cycles probe noncommuting string operators and cannot be combined,
so each captures only part of the logical information and none provides a
measure of magic.

In this paper we construct a single entanglement quantity, the
\emph{spinal reflected entropy}, that contains information from \emph{all} noncontractible cycles simultaneously
and reveals the magic content of the stored state. More explicitly, we first trace out a disk. A disk supports no logical operator and is therefore
correctable~\cite{knill1997theory}, so no logical information is lost. The remaining punctured surface
retracts onto a \emph{spine}~\cite{hatcher2002algebraic,matveev2007algorithmic}: $2g$ noncontractible loops meeting at a single
vertex, where $g$ is the genus of the original surface. We partition the remaining surface into a disk $A$ around the spine vertex and bands $B$ along its loops. The reduced state on $AB$ is mixed, and we evaluate its reflected entropy~\cite{dutta2021canonical}, a measure of correlations in mixed
states, across $A$ and $B$. Like the
entropy of a disk, it obeys an area law with a topological correction,
but also contains a new term:
\begin{equation}
 \begin{gathered}[b]
 \vcenter{\hbox{\includegraphics[width=.43\linewidth]{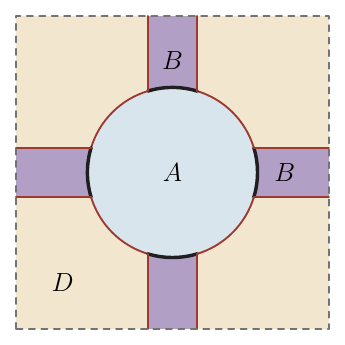}}}\\[4pt]
 \underbrace{S_R^{(n)}(A:B)}_{\mathclap{\substack{\text{spinal reflected}\\\text{entropy}}}}=
 \underbrace{\alpha_n L-4g\gamma}_{\mathclap{\substack{\text{independent of}\\\text{stored state}}}}
 \quad+\quad\underbrace{\zeta_n(\psi)}_{\mathclap{\substack{\text{depends on}\\\text{stored state}}}}.
 \end{gathered}
 \label{eq:intro-result}
\end{equation}
Here $n$ is the R\'enyi index. On the right-hand side, the first term is an
area-law term proportional to the boundary length $L$ with a nonuniversal coefficient $\alpha_n$, and the second is
the topological entanglement entropy, with one $\gamma$ for each of the $4g$
places where a band meets $A$. Both are the same for every ground state. The new term $\zeta_n(\psi)$, which we call the \emph{logical reflected
entropy}, is the part of the spinal reflected entropy that carries all the dependence on the stored state $\psi$. For the toric
code we find $\zeta_n(\psi)=M_n(\psi)+\log\mathrm{GSD}$, where $M_n(\psi)$ is the stabilizer
R\'enyi entropy, a widely used computable measure of magic~\cite{leone2022stabilizer}, of the encoded state $\psi$ and
$\mathrm{GSD}=4^g$ is the ground-state degeneracy. 
Since $M_n$ vanishes
exactly for stabilizer states, $\zeta_n$ takes its minimum value
$\log\mathrm{GSD}$ on stabilizer states, and any excess measures \emph{logical magic}.

This result has a topological origin: on the surface sewn to its reflected copy,
measuring the anyon charges threading the boundary between $A$ and $B$ amounts to Bell sampling of the stored
state~\cite{Montanaro:2017oht,Gross:2021ixc}, whose outcome probabilities are the squared
Pauli expectation values that define $M_n$. We explain this on
the torus, extend it to higher genus and to general stabilizer codes on qubits, and confirm it by an explicit lattice calculation in
the End Matter.

\begin{figure}[t]
\centering
\includegraphics[width=.98\linewidth]{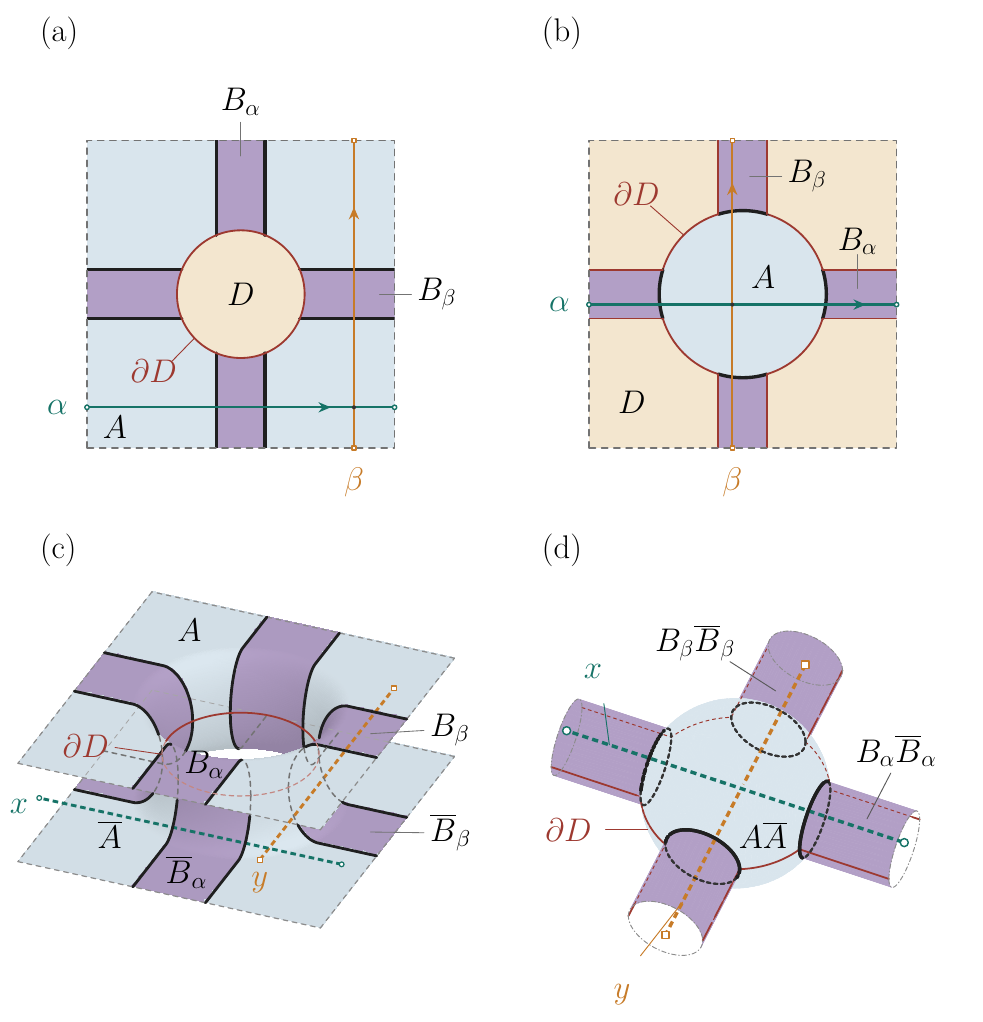}
\caption{The torus partition and the tube-charge basis.
(a) A planar fundamental domain of the torus, with opposite dashed edges identified.
The disk $D$ (beige) lies at the center; the blue corner pieces form the disk $A$, and
the purple regions form the bands $B_\alpha$ and $B_\beta$. The cycles $\alpha$ (teal)
and $\beta$ (orange) form a spine. Removing $D$ leaves $\Sigma_{1,1}=A\cup B$, with
$B=B_\alpha\cup B_\beta$ and boundary $\partial D$ (red).
(b) An equivalent partition obtained by translating the fundamental domain by half a
period in each direction and smoothly deforming the region boundaries.
(c),(d) The doubled surfaces obtained from (a) and (b), respectively, shown as the
boundary of the thickened punctured torus $\Sigma_{1,1}\times I$.
The original and reflected copies are sewn along $\partial D$ to form the genus-two
surface $\Sigma_2$. The path integral with an $x$ loop along $\alpha$
and a $y$ loop along $\beta$ inside the solid prepares $|x,y\rangle$ on this boundary.
In (d), $A\bar A$ is a four-holed sphere and the doubled bands form the two tubes
$B_\alpha\bar B_\alpha$ and $B_\beta\bar B_\beta$, drawn cut open.
The black interfaces between $A$ and $B$ in (a),(b) double to the four entanglement
circles $\partial(A\bar A)$ in (d), whose total length is $L$ in
Eq.~\eqref{eq:see-definition}.}
\label{fig:skeleton-partition-g1}
\end{figure}

\emph{Spinal reflected entropy for the toric code on a torus.---}
Consider the toric code on a lattice with periodic boundary conditions,
\begin{equation}
H=-\sum_v A_v-\sum_p B_p,\quad
A_v=\prod_{e\ni v}X_e,\quad B_p=\prod_{e\in\partial p}Z_e.
\label{eq:toric-hamiltonian}
\end{equation}
The four ground states satisfy $A_v=B_p=1$ and span the code space, which stores two logical qubits.
Physical $Z$ strings transport electric charges $e$, and physical $X$
strings on the dual lattice transport magnetic fluxes
$m$; their composite is the fermion $f=e\times m$. Let $\alpha$ and $\beta$ be two
noncontractible cycles of the torus that intersect once
[Fig.~\ref{fig:skeleton-partition-g1}(a,b)]. Choose
direct-lattice representatives of $\alpha,\beta$ and parallel dual-lattice
representatives $\alpha^*,\beta^*$. The logical Pauli operators are
$Z_1=\prod_{e\in\alpha}Z_e$,
$X_1=\prod_{e\,\mathrm{crossed\ by}\,\beta^*}X_e$,
$Z_2=\prod_{e\in\beta}Z_e$, and
$X_2=\prod_{e\,\mathrm{crossed\ by}\,\alpha^*}X_e$.
More generally, the string operator that moves an anyon $x$ around $\alpha$ and an
anyon $y$ around $\beta$ is a logical Pauli operator. For $x=e^am^b$ and $y=e^cm^d$,
with exponents in $\{0,1\}$,
$P_{xy}\propto(Z_1^aX_2^b)(Z_2^cX_1^d)$,
with the phase chosen such that $P_{xy}$ is Hermitian. The sixteen pairs $(x,y)$, with
$x,y\in\{1,e,m,f\}$, thus label the sixteen two-qubit Pauli strings.

Throughout this paper, we denote by $\Sigma_{g,b}$ a compact oriented
surface of genus $g$ with $b$ boundary components, and write
$\Sigma_g=\Sigma_{g,0}$ for the closed surface; for instance, the
torus is $T^2=\Sigma_1$. In this work, we consider pure logical states
$|\psi\rangle$ in the ground-state manifold of the Hamiltonian~\eqref{eq:toric-hamiltonian} on a closed surface.
For such a state on $\Sigma_1$, we now define the
spinal reflected entropy explicitly and extract its logical piece in three steps
(Fig.~\ref{fig:skeleton-partition-g1}).
\begin{enumerate}
\item \emph{Trace out and vectorize.} Trace out a disk
$D\subset\Sigma_1$, assigning to the punctured torus
$\Sigma_{1,1}=\Sigma_1\setminus D$ the reduced density matrix
$\rho\coloneqq\Tr_D|\psi\rangle\langle\psi|$. The figures draw $D$ large or small as convenient, since a disk is correctable regardless of its size [Fig.~\ref{fig:skeleton-partition-g1}(a,b)]. Then introduce a reflected
copy $\bar\Sigma_{1,1}$ and vectorize $\rho$:
\begin{equation}
|\Omega_\psi\rangle\coloneqq\frac{1}{\sqrt{\Tr\rho^2}}
\sum_{i,j}\langle i|\rho|j\rangle\,
|i\rangle_{\Sigma_{1,1}}|\bar j\rangle_{\bar\Sigma_{1,1}},
\label{eq:toric-vectorization}
\end{equation}
where $\{|i\rangle\}$ is an orthonormal basis and $|\bar i\rangle$ its
complex conjugate.

Vectorization turns the mixed state $\rho$ into a
pure state, also known as its Choi state. It is the canonical purification of
$\rho^2/\Tr\rho^2$. For the toric code $\rho$ is
proportional to a projector (see End Matter), so $\rho^2/\Tr\rho^2=\rho$ and
$|\Omega_\psi\rangle$ is
also the canonical purification of $\rho$ itself.
Topologically, $|\Omega_\psi\rangle$ lives on the double of the
punctured torus: sewing $\Sigma_{1,1}$ and $\bar\Sigma_{1,1}$ along
$\partial D$ gives a closed surface of genus two, $\Sigma_2$
[Fig.~\ref{fig:skeleton-partition-g1}(c,d)]. The single-copy problem on
$\Sigma_1$ thus becomes a pure-state problem on $\Sigma_2$.

\item \emph{Partition along the spine.} Partition $\Sigma_{1,1}$
into the disk $A$ around the vertex and the two bands $B_\alpha$ and $B_\beta$
along $\alpha$ and $\beta$, with $B=B_\alpha\cup B_\beta$
[Fig.~\ref{fig:skeleton-partition-g1}(a,b)]. We partition the reflected copy into $\bar A\cup\bar B$ accordingly, and set
\begin{equation}
\rho_{A\bar A}\coloneqq\Tr_{B\bar B}|\Omega_\psi\rangle\langle\Omega_\psi|.
\label{eq:torus-partition}
\end{equation}
The R\'enyi entropy $S_n(\rho_{A\bar A})$ is the reflected entropy
$S_R^{(n)}(A:B)$ of $\rho$; we
call it the \emph{spinal reflected entropy}.
On the doubled surface, the partition takes a simple shape
[Fig.~\ref{fig:skeleton-partition-g1}(d)]. Each band, joined to its reflected copy along
the two sides that lie on $\partial D$, becomes a tube, and the vertex disk, joined to
its reflected copy, becomes a sphere with four holes, one for each end of a tube. Thus
$A\bar A$ is a four-holed sphere and $B\bar B$ is the disjoint union of the two
tubes $B_\alpha\bar B_\alpha$ and $B_\beta\bar B_\beta$.
Equivalently, $S_R^{(n)}(A:B)$ is the
operator entanglement of $\rho$ across
$A:B$~\cite{zanardi2001entanglement,prosen2007operator,Dubail:2016xht}.
Reflected entropy of topological phases has been studied for adjacent
regions~\cite{Berthiere:2020ihq,Sohal:2023hst,Liu:2023pdz}, where its
universal part detects the anyon content. Here the bands run along all
generating cycles at once, and this is what lets the entropy resolve all
logical operators jointly.
\item \emph{Extract the logical piece.} For $g=1$,
Eq.~\eqref{eq:intro-result} reads
\begin{equation}
S_R^{(n)}(A:B)=\alpha_n L-4\gamma+\zeta_n(\psi),
\label{eq:see-definition}
\end{equation}
with $\gamma=\log\mathcal D=\log 2$ for the toric code. On the double, the
cut $\partial(A\bar A)$ consists of four entanglement circles, the rims of the four
holes, and $L$ is their total length [Fig.~\ref{fig:skeleton-partition-g1}(d)]. Each circle
carries the topological entanglement entropy~\cite{Hamma:2004vdz}, the value
it would have if no anyon flux threaded it; this contribution is the same
for every ground state and every $n$~\cite{Flammia:2009axf}. The remaining constant $\zeta_n(\psi)$ is the logical reflected entropy.
\end{enumerate}

\emph{Topological explanation of the logical magic.---}
Having set up the spinal reflected entropy in Eq.~\eqref{eq:see-definition}, we now
show that the constant $\zeta_n(\psi)$ captures the logical magic. What we need is the
Schmidt decomposition of $|\Omega_\psi\rangle$ across the cut between $A\bar A$ and
$B\bar B$. We obtain it in three steps: (i) we express $|\Omega_\psi\rangle$ in a basis
labeled by the logical Pauli operators; (ii) we show that each basis state carries a
definite anyon charge through each of the two tubes; (iii) we read off the spectrum of
$\rho_{A\bar A}$.

\emph{(i) The state.} A logical pure state is completely specified by the expectation
values of the logical Pauli operators,
\begin{equation}
|\psi\rangle\langle\psi|=\sum_{x,y}\langle\psi|P_{xy}|\psi\rangle\,P_{xy}\,\frac{\Pi}{4},
\label{eq:psi-pauli}
\end{equation}
where $\Pi$ is the projector onto the code space, so that $\Pi/4$ is the maximally
mixed logical state. Choose the strings $P_{xy}$ away from $D$. The partial trace over
$D$ then acts on $\Pi$ alone, and
\begin{equation}
\rho=\sum_{x,y}\langle\psi|P_{xy}|\psi\rangle\,P_{xy}\,\rho_0,\qquad
\rho_0\coloneqq\Tr_D\frac{\Pi}{4}.
\label{eq:rho-pauli}
\end{equation}
Here $\rho_0$ is the reduced density matrix of the maximally mixed logical state; it
knows nothing about $\psi$. We now vectorize as in Eq.~\eqref{eq:toric-vectorization},
where an operator multiplying $\rho$ from the left acts on the first copy, and obtain
\begin{equation}
|\Omega_\psi\rangle=\frac12\sum_{x,y}\langle\psi|P_{xy}|\psi\rangle\,|x,y\rangle,\qquad
|x,y\rangle\coloneqq(P_{xy}\otimes\mathbb{I})|\Omega_0\rangle,
\label{eq:omega-pauli}
\end{equation}
where $|\Omega_0\rangle$ is the normalized vectorization of $\rho_0$. The sixteen states
$|x,y\rangle$ are orthonormal, because $\rho_0$ is proportional to a projector and distinct
logical Pauli operators are orthogonal on the code space. The prefactor follows from
$\sum_{x,y}|\langle\psi|P_{xy}|\psi\rangle|^2=4$, which holds for every pure state of two
qubits. Notice that in the Choi state~(\ref{eq:omega-pauli}), all information on the logical state $|\psi\rangle$ is encoded in the expectation values of the logical Pauli strings. In particular, the basis $|x,y\rangle$ is independent of the choice of $|\psi\rangle$.

\emph{(ii) The basis.}
The physical meaning of the basis states $|x,y\rangle$ becomes particularly transparent in the path-integral picture.
In a topological phase, the path integral over a three-dimensional region prepares a ground
state on the surface that bounds it, and anyon lines inserted in the region select which
ground state~\cite{atiyah1988topological,liu2024anyon}. 
For the toric code on a torus boundary, the path integral is naturally defined on a solid torus.
Left empty, it
prepares the ground state with no anyon flux through the disk bounded by $\beta$; an anyon
line $e^am^b$ along its core, acting on the surface as $Z_1^aX_2^b$, prepares the other three.

The operator $\rho_0$ has such a description too. Up to normalization, it is the
projector onto the states of $\Sigma_{1,1}$ that satisfy every stabilizer constraint inside
$\Sigma_{1,1}$, so it is prepared by the path integral on the slab $\Sigma_{1,1}\times I$,
the punctured torus thickened by an interval $I$: the top face carries the ket and the
bottom face the bra. 
Multiplying $\rho_0$ by $P_{xy}$ inserts an $x$ loop along $\alpha$
and a $y$ loop along $\beta$ into the slab [Fig.~\ref{fig:skeleton-partition-g1}(c)]. Vectorization
reads the slab as a state on its whole boundary. The two faces, joined along $\partial D$,
form the doubled surface $\Sigma_2$, so $|x,y\rangle$ is the state on $\Sigma_2$ prepared by
the thickened punctured torus with the two anyon loops inside.

In Fig.~\ref{fig:skeleton-partition-g1}(d) the same solid is drawn as a ball with two handles: the
tubes are the walls of the handles, the $x$ loop threads the first, and the $y$ loop the
second. The anyon charge passing through a tube is measured by a closed string operator that
encircles the tube. In the solid, this circle bounds a disk across the handle. If nothing
pierces the disk, the string can be contracted through the solid and acts trivially; if the
anyon line $x$ pierces it, the string registers $x$ through the braiding phase. Hence
$|x,y\rangle$ carries charge $x$ through the first tube and charge $y$ through the second,
and Eq.~\eqref{eq:omega-pauli} is precisely the expansion of $|\Omega_\psi\rangle$ in the
\emph{tube-charge basis}. The joint tube charge $(x,y)$ occurs with probability
\begin{equation}
p_{xy}=\frac14|\langle\psi|P_{xy}|\psi\rangle|^2.
\label{eq:tube-probability}
\end{equation}

The same conclusion that the amplitudes in the tube-charge basis describe the logical Pauli expectation values follows from an independent path-integral derivation and explicit lattice calculation; see the End Matter.

\emph{(iii) The spectrum.} The Schmidt decomposition is now immediate. A string that
encircles a tube can be placed on the tube itself or pushed across its rim onto the sphere,
so the tube charges can be measured within $B\bar B$ and also within $A\bar A$. States with
different tube charges are therefore orthogonal on either side of the cut, and tracing out
$B\bar B$ leaves no cross terms:
\begin{equation}
\rho_{A\bar A}=\sum_{x,y}p_{xy}\,\rho^{(xy)}_{A\bar A},
\label{eq:tube-blocks}
\end{equation}
where $\rho^{(xy)}_{A\bar A}$ is the reduced state of $|x,y\rangle$ on $A\bar A$ and the sixteen
blocks have mutually orthogonal supports. The blocks differ only by a unitary on $A$, the
part of the string $P_{xy}$ inside $A$, so they share the spectrum of $\rho^{(11)}_{A\bar A}$, and
$S_n(\rho_{A\bar A})=S_n(\rho^{(11)}_{A\bar A})+H_n(p)$, where
$H_n(p)=(1-n)^{-1}\log\sum_{x,y}p_{xy}^{\,n}$ is the R\'enyi entropy of the distribution
$p$. The trivial block carries no flux through the entanglement circles, so
$S_n(\rho^{(11)}_{A\bar A})=\alpha_nL-4\gamma$. Comparing with Eq.~\eqref{eq:see-definition} gives
$\zeta_n=H_n(p)$: the logical reflected entropy is the uncertainty of the joint tube
charge.

The distribution $p$ is familiar from the study of magic: it defines the
stabilizer R\'enyi entropy, which we now recall. For a pure state
$|\psi\rangle$ of $N$ qubits, the squared Pauli expectation values define
a probability distribution over the $4^N$ Pauli strings $P_i$,
\begin{equation}
p_i=\frac{|\langle\psi|P_i|\psi\rangle|^2}{2^N},\qquad \sum_i p_i=1,
\label{eq:pauli-distribution}
\end{equation}
where the normalization follows from purity. A stabilizer state
spreads $p$ uniformly over the $2^N$ elements of its stabilizer group, and
magic spreads it further. The stabilizer R\'enyi entropy measures this
excess,
\begin{equation}
M_n(\psi)=H_n(p)-\log 2^N.
\label{eq:sre}
\end{equation}
Since every $p_i\le 1/2^N$, $M_n\ge0$, with equality exactly for
stabilizer states. For $n\ge2$, $M_n$ is
moreover a magic
monotone~\cite{Emerson:2013zse,Leone:2024lfr,Haug:2023hcs} under deterministic stabilizer operations whose inputs and outputs are pure.

For the toric code on the torus, $N=2$, the Pauli strings are the $P_{xy}$ defined above,
and $p_i$ is the tube-charge distribution $p_{xy}$ of
Eq.~\eqref{eq:tube-probability}. Since $2^N=4$ is the ground-state degeneracy, we immediately identify
\begin{equation}
\zeta_n(\psi)=M_n(\psi)+\log\mathrm{GSD}.
\label{eq:torus-sre}
\end{equation}
The ground-state degeneracy sets the floor $\zeta_n\ge\log\mathrm{GSD}$,
reached exactly by logical stabilizer states, and the excess is the logical
magic.

\begin{figure}[!t]
\centering
\includegraphics[width=.98\linewidth]{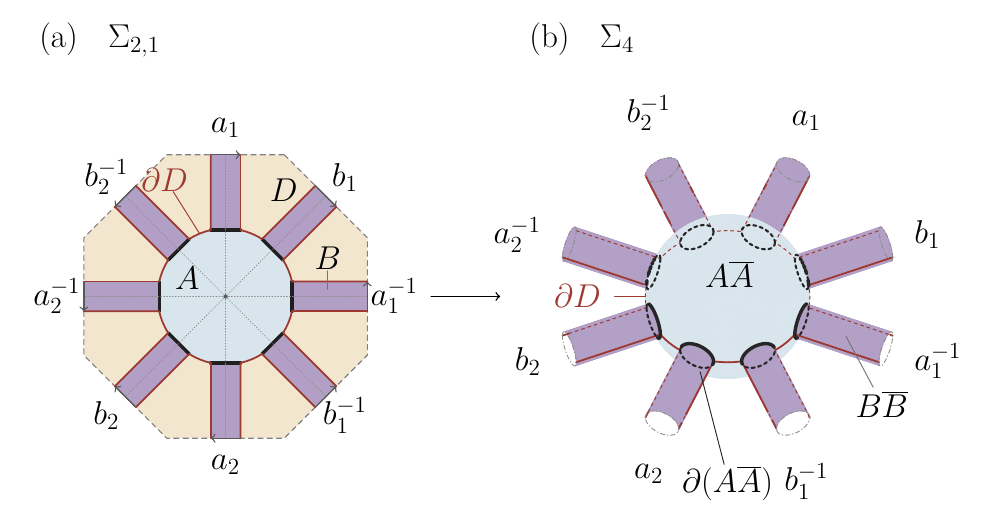}
\caption{The spinal partition for a genus-two surface.
(a) The surface drawn as an octagon whose sides are identified in pairs, $a_i$ with
$a_i^{-1}$ and $b_i$ with $b_i^{-1}$. The disk $A$ (blue) at the center and the four bands
$B$ (purple), each passing through a pair of identified sides, form a neighborhood of the
spine. The corners (beige) join into the single disk $D$ that is traced out, with boundary
$\partial D$ (red).
(b) Sewing to the reflected copy along $\partial D$ gives $\Sigma_4$. The region $A\bar A$
is a sphere with eight holes, and the bands become four tubes, drawn cut open with their
ends identified as labeled. The eight black circles form the entanglement cut.}
\label{fig:skeleton-partition}
\end{figure}

\emph{Higher genus and general stabilizer codes.---}
The above result generalizes directly to the toric code on a surface of higher genus.
On a closed surface $\Sigma_g$, choose $2g$ loops $a_1,b_1,\dots,a_g,b_g$ through a
common vertex that generate all noncontractible cycles; they form a spine. Its neighborhood
consists of a disk $A$ around the vertex and $2g$ bands $B$, one along each loop, and the
complement is a single disk $D$. We trace out $D$, vectorize, and partition exactly as on
the torus. Sewing the punctured surface $\Sigma_{g,1}=\Sigma_g\setminus D$ to its reflected
copy along $\partial D$ gives a closed surface of genus $2g$, on which $A\bar A$ is a sphere
with $4g$ holes and $B\bar B$ consists of $2g$ tubes. Figure~\ref{fig:skeleton-partition}
shows a genus-two example. The spinal reflected entropy takes the form of
Eq.~\eqref{eq:intro-result}, with one $\gamma$ for each of the $4g$ entanglement circles.

The toric code on $\Sigma_g$ encodes $2g$ logical qubits, and the argument carries over
step by step. The logical Pauli strings $P\in\calP_{2g}=\{I,X,Y,Z\}^{\otimes 2g}$ are
products of string operators along the $2g$ loops. Equation~\eqref{eq:omega-pauli} holds
with the sum running over $\calP_{2g}$, the prefactor $2^{-g}$, and the basis states
$|P\rangle\coloneqq(P\otimes\mathbb{I})|\Omega_0\rangle$. The thickened punctured surface
$\Sigma_{g,1}\times I$ is a ball with $2g$ handles whose walls are the $2g$ tubes, and the
anyon that $P$ moves around a loop threads the corresponding handle. Each $|P\rangle$
therefore has definite charges through all the tubes, the blocks of $\rho_{A\bar A}$ are
again orthogonal with identical spectra, and
\begin{equation}
\zeta_n(\psi)=H_n(p),\qquad
p_P=\frac{|\langle\psi|P|\psi\rangle|^2}{4^g},\quad P\in\calP_{2g}.
\label{eq:toric-higher-genus}
\end{equation}
Since $\mathrm{GSD}=4^g$, Eq.~\eqref{eq:torus-sre} generalizes to
\begin{equation}
\zeta_n(\psi)=M_n(\psi)+\log\mathrm{GSD}
=M_n(\psi)+2g\log2.
\label{eq:logical-stabilizer-entropy}
\end{equation}
The logical reflected entropy is independent of the choice of spine, since a
different set of generating cycles only relabels the logical Pauli operators.

The same structure appears in general two-dimensional topological stabilizer codes
on qubits.
With translation invariance the statement is microscopic: translation-invariant
two-dimensional topological Pauli stabilizer codes on qubits decompose, after
coarse-graining, into a finite number $r$ of toric-code layers and decoupled
qubits~\cite{Haah:2018classification}, with an explicit construction for CSS
codes~\cite{Wang:2026decoupling}. 
Toric-code anyons are expected even without translation invariance. If every anyon is
transported by Pauli string operators, all topological spins are $\pm1$, and the anyon
theory is then that of a stack of toric-code layers~\footnote{We assume a homogeneous
bulk with a modular anyon theory. Exchanging two identical anyons that are moved by Pauli
strings gives only a sign. An anyon theory with all spins $\pm1$ decomposes into toric-code
and three-fermion factors~\cite{Balasubramanian:2024bfb}. A strictly two-dimensional local
commuting-projector Hamiltonian has vanishing thermal Hall
conductance~\cite{Kapustin:2019thermal}, whereas each three-fermion factor contributes
chiral central charge $4$ modulo $8$. These factors therefore occur in pairs, and each
pair is equivalent, as an anyon theory, to two toric-code layers.}.

The logical reflected entropy follows from the anyon theory alone. On a closed,
untwisted surface $\Sigma_g$, $r$ layers encode $k=2gr$ logical qubits, and $\gamma=r\log2$
in Eq.~\eqref{eq:intro-result}. The joint charges through the $2g$ tubes label all $4^k$
logical Pauli strings, so that
\begin{equation}
\zeta_n(\psi)=M_n(\psi)+k\log2,
\label{eq:general-stabilizer}
\end{equation}
even for states entangled across the layers. The lattice calculation in the End Matter
uses only the stabilizer structure, and it gives the same identity directly whenever $A$,
the union $B$ of all the bands, and $D$ are each correctable.

\emph{Discussion.---}
Entanglement has long been used to tell which topological phase a system
is in. We have shown that entanglement can also reveal the logical magic stored in the state: the Rényi uncertainty of the anyon charges threading the tubes of the sewn surface equals the stabilizer Rényi entropy plus \(\log \mathrm{GSD}\), 
this links the study of magic in many-body
systems~\cite{Liu:2020yso,White:2020zoz,Nehra:2025tha,Zhang:2024fyp,Zhang:2025dhg,zhang2026extensive}
to the entanglement structure of topological order.

So far we have worked with qubits, where the Pauli group and the stabilizer R\'enyi
entropy take their standard form. For qudits of dimension $N$, the stabilizer R\'enyi
entropy is defined through the generalized Pauli operators, also known as Weyl operators,
and our results extend to the $\ZZ_N$ toric code. The string operators of the $N^2$ anyons
$e^am^b$ are logical Weyl operators, the tube charges are distributed as the squared Weyl
expectation values of the stored state, and
$\zeta_n=M_n+\log\mathrm{GSD}$ with $\mathrm{GSD}=N^{2g}$. We do not claim such a relation
for general stabilizer codes on $\ZZ_N$ qudits. For composite $N$ new phases appear, for
example the double semion on $\ZZ_4$ qudits~\cite{Ellison:2021vth}, and a decomposition
into toric-code layers is no longer available.

\emph{Acknowledgments.---}
We thank Hong-Yu Wang, Milo Moses, Bohao Zhuang and Meng Sun
for discussions.
This work is supported by the National Natural Science Foundation of China under Grant No.~12575022 and No. 12375027, the Scientific Research Innovation Capability Support Project for Young Faculty under Grant No.~ZY2025157, the Shanghai Committee of Science and Technology under Grant No.~25LZ2600800, and the Tsinghua Dushi Program.

\nocite{Balasubramanian:2024bfb,Kapustin:2019thermal}
\bibliography{ref}

@article{kitaev2003fault,
  author        = {Kitaev, A. {\relax Yu}.},
  title         = {Fault-tolerant quantum computation by anyons},
  journal       = {Annals Phys.},
  volume        = {303},
  number        = {1},
  pages         = {2--30},
  year          = {2003},
  doi           = {10.1016/S0003-4916(02)00018-0},
  eprint        = {quant-ph/9707021},
  archivePrefix = {arXiv}
}

@article{atiyah1988topological,
  author        = {Atiyah, Michael F.},
  title         = {Topological quantum field theories},
  journal       = {Publ. Math. IHES},
  volume        = {68},
  pages         = {175--186},
  year          = {1988},
  doi           = {10.1007/BF02698547}
}

@article{knill1997theory,
  author        = {Knill, Emanuel and Laflamme, Raymond},
  title         = {Theory of quantum error-correcting codes},
  journal       = {Phys. Rev. A},
  volume        = {55},
  number        = {2},
  pages         = {900--911},
  year          = {1997},
  doi           = {10.1103/PhysRevA.55.900},
  eprint        = {quant-ph/9604034},
  archivePrefix = {arXiv}
}

@article{kitaev2006topological,
  author        = {Kitaev, Alexei and Preskill, John},
  title         = {Topological entanglement entropy},
  journal       = {Phys. Rev. Lett.},
  volume        = {96},
  number        = {11},
  pages         = {110404},
  year          = {2006},
  doi           = {10.1103/PhysRevLett.96.110404},
  eprint        = {hep-th/0510092},
  archivePrefix = {arXiv}
}

@article{levin2006detecting,
  author        = {Levin, Michael and Wen, Xiao-Gang},
  title         = {Detecting topological order in a ground state wave function},
  journal       = {Phys. Rev. Lett.},
  volume        = {96},
  number        = {11},
  pages         = {110405},
  year          = {2006},
  doi           = {10.1103/PhysRevLett.96.110405},
  eprint        = {cond-mat/0510613},
  archivePrefix = {arXiv}
}

@article{dong2008topological,
  author        = {Dong, Shiying and Fradkin, Eduardo and Leigh, Robert G.
                   and Nowling, Sean},
  title         = {Topological entanglement entropy in {Chern--Simons} theories
                   and quantum {Hall} fluids},
  journal       = {JHEP},
  volume        = {2008},
  number        = {05},
  pages         = {016},
  year          = {2008},
  doi           = {10.1088/1126-6708/2008/05/016},
  eprint        = {0802.3231},
  archivePrefix = {arXiv},
  primaryClass  = {hep-th}
}

@article{bonderson2017anyonic,
  author        = {Bonderson, Parsa and Knapp, Christina and Patel, Kaushal},
  title         = {Anyonic entanglement and topological entanglement entropy},
  journal       = {Annals Phys.},
  volume        = {385},
  pages         = {399--468},
  year          = {2017},
  doi           = {10.1016/j.aop.2017.07.018},
  eprint        = {1706.09420},
  archivePrefix = {arXiv},
  primaryClass  = {quant-ph}
}

@article{leone2022stabilizer,
  author        = {Leone, Lorenzo and Oliviero, Salvatore F. E. and Hamma, Alioscia},
  title         = {Stabilizer {R\'enyi} entropy},
  journal       = {Phys. Rev. Lett.},
  volume        = {128},
  number        = {5},
  pages         = {050402},
  year          = {2022},
  doi           = {10.1103/PhysRevLett.128.050402},
  eprint        = {2106.12587},
  archivePrefix = {arXiv},
  primaryClass  = {quant-ph}
}

@article{liu2024anyon,
  author  = {Liu, Shang},
  title   = {Anyon quantum dimensions from an arbitrary ground state wave function},
  journal = {Nat. Commun.},
  volume  = {15},
  pages   = {5134},
  year    = {2024},
  doi     = {10.1038/s41467-024-47856-7}
}

@book{hatcher2002algebraic,
  author        = {Hatcher, Allen},
  title         = {Algebraic Topology},
  publisher     = {Cambridge University Press},
  address       = {Cambridge},
  year          = {2002},
  isbn          = {978-0-521-79540-1},
  url           = {https://pi.math.cornell.edu/~hatcher/AT/ATpage.html}
}

@article{zhang2012quasiparticle,
  author        = {Zhang, Yi and Grover, Tarun and Turner, Ari and
                   Oshikawa, Masaki and Vishwanath, Ashvin},
  title         = {Quasiparticle statistics and braiding from ground-state entanglement},
  journal       = {Phys. Rev. B},
  volume        = {85},
  pages         = {235151},
  year          = {2012},
  doi           = {10.1103/PhysRevB.85.235151},
  eprint        = {1111.2342},
  archivePrefix = {arXiv},
  primaryClass  = {cond-mat.str-el}
}

@article{zanardi2001entanglement,
  author        = {Zanardi, Paolo},
  title         = {Entanglement of quantum evolutions},
  journal       = {Phys. Rev. A},
  volume        = {63},
  pages         = {040304},
  year          = {2001},
  doi           = {10.1103/PhysRevA.63.040304},
  eprint        = {quant-ph/0010074},
  archivePrefix = {arXiv}
}

@article{prosen2007operator,
  author        = {Prosen, Toma{\v z} and Pi{\v z}orn, Iztok},
  title         = {Operator space entanglement entropy in a transverse {Ising} chain},
  journal       = {Phys. Rev. A},
  volume        = {76},
  pages         = {032316},
  year          = {2007},
  doi           = {10.1103/PhysRevA.76.032316},
  eprint        = {0706.2480},
  archivePrefix = {arXiv},
  primaryClass  = {quant-ph}
}

@article{dutta2021canonical,
  author        = {Dutta, Souvik and Faulkner, Thomas},
  title         = {A canonical purification for the entanglement wedge cross-section},
  journal       = {JHEP},
  volume        = {2021},
  number        = {03},
  pages         = {178},
  year          = {2021},
  doi           = {10.1007/JHEP03(2021)178},
  eprint        = {1905.00577},
  archivePrefix = {arXiv},
  primaryClass  = {hep-th}
}

@misc{zhang2026extensive,
  author        = {Zhang, Yuzhen and Kim, Isaac H. and Bao, Yimu and Vijay, Sagar},
  title         = {Extensive long-range magic in non-{Abelian} topological orders},
  year          = {2026},
  eprint        = {2605.15150},
  archivePrefix = {arXiv},
  primaryClass  = {quant-ph}
}

@article{dijkgraaf1990gauge,
  author = {Dijkgraaf, Robbert and Witten, Edward},
  title = {Topological gauge theories and group cohomology},
  journal = {Communications in Mathematical Physics},
  volume = {129},
  pages = {393--429},
  year = {1990},
  doi = {10.1007/BF02096988}
}

@article{Berthiere:2020ihq,
    author = "Berthiere, Cl{\'e}ment and Chen, Hongjie and Liu, Yuefeng and Chen, Bin",
    title = "{Topological reflected entropy in Chern-Simons theories}",
    eprint = "2008.07950",
    archivePrefix = "arXiv",
    primaryClass = "hep-th",
    doi = "10.1103/PhysRevB.103.035149",
    journal = "Phys. Rev. B",
    volume = "103",
    number = "3",
    pages = "035149",
    year = "2021"
}

@article{Sohal:2023hst,
    author = "Sohal, Ramanjit and Ryu, Shinsei",
    title = "{Entanglement in tripartitions of topological orders: A diagrammatic approach}",
    eprint = "2301.07763",
    archivePrefix = "arXiv",
    primaryClass = "cond-mat.str-el",
    doi = "10.1103/PhysRevB.108.045104",
    journal = "Phys. Rev. B",
    volume = "108",
    number = "4",
    pages = "045104",
    year = "2023"
}

@article{Liu:2023pdz,
    author = "Liu, Yuhan and Kusuki, Yuya and Kudler-Flam, Jonah and Sohal, Ramanjit and Ryu, Shinsei",
    title = "{Multipartite entanglement in two-dimensional chiral topological liquids}",
    eprint = "2301.07130",
    archivePrefix = "arXiv",
    primaryClass = "cond-mat.str-el",
    doi = "10.1103/PhysRevB.109.085108",
    journal = "Phys. Rev. B",
    volume = "109",
    number = "8",
    pages = "085108",
    year = "2024"
}

@article{Bravyi:2004isx,
    author = "Bravyi, Sergei and Kitaev, Alexei",
    title = "{Universal quantum computation with ideal Clifford gates and noisy ancillas}",
    eprint = "quant-ph/0403025",
    archivePrefix = "arXiv",
    doi = "10.1103/PhysRevA.71.022316",
    journal = "Phys. Rev. A",
    volume = "71",
    number = "2",
    pages = "022316",
    year = "2005"
}

@article{Emerson:2013zse,
    author = "Veitch, Victor and Hamed Mousavian, Seyed Ali and Gottesman, Daniel and Emerson, Joseph",
    title = "{The resource theory of stabilizer quantum computation}",
    eprint = "1307.7171",
    archivePrefix = "arXiv",
    primaryClass = "quant-ph",
    doi = "10.1088/1367-2630/16/1/013009",
    journal = "New J. Phys.",
    volume = "16",
    number = "1",
    pages = "013009",
    year = "2014"
}

@inproceedings{Gottesman:1998hu,
    author = "Gottesman, Daniel",
    title = "{The Heisenberg representation of quantum computers}",
    booktitle = "{Group22: Proceedings of the XXII International Colloquium on Group Theoretical Methods in Physics}",
    eprint = "quant-ph/9807006",
    archivePrefix = "arXiv",
    reportNumber = "LAUR-98-2848, LA-UR-98-2848",
    pages = "32--43",
    editor = "Corney, S. P. and Delbourgo, R. and Jarvis, P. D.",
    publisher = "International Press",
    address = "Cambridge, MA",
    year = "1999"
}

@article{Leone:2024lfr,
    author = "Leone, Lorenzo and Bittel, Lennart",
    title = "{Stabilizer entropies are monotones for magic-state resource theory}",
    eprint = "2404.11652",
    archivePrefix = "arXiv",
    primaryClass = "quant-ph",
    doi = "10.1103/PhysRevA.110.L040403",
    journal = "Phys. Rev. A",
    volume = "110",
    number = "4",
    pages = "L040403",
    year = "2024"
}

@article{Haug:2023hcs,
    author = "Haug, Tobias and Piroli, Lorenzo",
    title = "{Stabilizer entropies and nonstabilizerness monotones}",
    eprint = "2303.10152",
    archivePrefix = "arXiv",
    primaryClass = "quant-ph",
    doi = "10.22331/q-2023-08-28-1092",
    journal = "Quantum",
    volume = "7",
    pages = "1092",
    year = "2023"
}

@article{Gross:2021ixc,
    author = "Gross, David and Nezami, Sepehr and Walter, Michael",
    title = "{Schur{\textendash}Weyl Duality for the Clifford Group with Applications: Property Testing, a Robust Hudson Theorem, and de Finetti Representations}",
    eprint = "1712.08628",
    archivePrefix = "arXiv",
    primaryClass = "quant-ph",
    doi = "10.1007/s00220-021-04118-7",
    journal = "Commun. Math. Phys.",
    volume = "385",
    number = "3",
    pages = "1325--1393",
    year = "2021"
}

@misc{Montanaro:2017oht,
    author = "Montanaro, Ashley",
    title = "{Learning stabilizer states by Bell sampling}",
    eprint = "1707.04012",
    archivePrefix = "arXiv",
    primaryClass = "quant-ph",
    month = "7",
    year = "2017"
}

@article{Liu:2020yso,
    author = "Liu, Zi-Wen and Winter, Andreas",
    title = "{Many-Body Quantum Magic}",
    eprint = "2010.13817",
    archivePrefix = "arXiv",
    primaryClass = "quant-ph",
    doi = "10.1103/PRXQuantum.3.020333",
    journal = "PRX Quantum",
    volume = "3",
    number = "2",
    pages = "020333",
    year = "2022"
}

@article{White:2020zoz,
    author = "White, Christopher David and Cao, ChunJun and Swingle, Brian",
    title = "{Conformal field theories are magical}",
    eprint = "2007.01303",
    archivePrefix = "arXiv",
    primaryClass = "quant-ph",
    doi = "10.1103/PhysRevB.103.075145",
    journal = "Phys. Rev. B",
    volume = "103",
    number = "7",
    pages = "075145",
    year = "2021"
}

@misc{Nehra:2025tha,
    author = "Nehra, Ritu and Tarabunga, Poetri Sonya and Frau, Martina and Collura, Mario and Tirrito, Emanuele and Dalmonte, Marcello",
    title = "{Topological magic response in quantum spin chains}",
    eprint = "2512.16673",
    archivePrefix = "arXiv",
    primaryClass = "quant-ph",
    month = "12",
    year = "2025"
}

@article{Dennis:2001nw,
    author = "Dennis, Eric and Kitaev, Alexei and Landahl, Andrew and Preskill, John",
    title = "{Topological quantum memory}",
    eprint = "quant-ph/0110143",
    archivePrefix = "arXiv",
    reportNumber = "CALT-68-2346, CALT-68-2346",
    doi = "10.1063/1.1499754",
    journal = "J. Math. Phys.",
    volume = "43",
    pages = "4452--4505",
    year = "2002"
}

@article{Hamma:2004vdz,
    author = "Hamma, Alioscia and Ionicioiu, Radu and Zanardi, Paolo",
    title = "{Bipartite entanglement and entropic boundary law in lattice spin systems}",
    eprint = "quant-ph/0409073",
    archivePrefix = "arXiv",
    doi = "10.1103/PhysRevA.71.022315",
    journal = "Phys. Rev. A",
    volume = "71",
    pages = "022315",
    year = "2005"
}

@article{Flammia:2009axf,
    author = "Flammia, Steven T. and Hamma, Alioscia and Hughes, Taylor L. and Wen, Xiao-Gang",
    title = "{Topological Entanglement R\'enyi Entropy and Reduced Density Matrix Structure}",
    eprint = "0909.3305",
    archivePrefix = "arXiv",
    primaryClass = "cond-mat.str-el",
    doi = "10.1103/PhysRevLett.103.261601",
    journal = "Phys. Rev. Lett.",
    volume = "103",
    pages = "261601",
    year = "2009"
}

@article{Dubail:2016xht,
    author = "Dubail, J.",
    title = "{Entanglement scaling of operators: a conformal field theory approach, with a glimpse of simulability of long-time dynamics in 1  +  1d}",
    eprint = "1612.08630",
    archivePrefix = "arXiv",
    primaryClass = "cond-mat.str-el",
    doi = "10.1088/1751-8121/aa6f38",
    journal = "J. Phys. A",
    volume = "50",
    number = "23",
    pages = "234001",
    year = "2017"
}

@article{Zhang:2024fyp,
    author = "Zhang, Yuzhen and Gu, Yingfei",
    title = "{Quantum magic dynamics in random circuits}",
    eprint = "2410.21128",
    archivePrefix = "arXiv",
    primaryClass = "quant-ph",
    doi = "10.1038/s41534-026-01253-8",
    journal = "npj Quantum Inf.",
    volume = "12",
    number = "1",
    pages = "87",
    year = "2026"
}

@article{Zhang:2025dhg,
    author = "Zhang, Yuzhen and Vijay, Sagar and Gu, Yingfei and Bao, Yimu",
    title = "{Designs from Magic-Augmented Clifford Circuits}",
    eprint = "2507.02828",
    archivePrefix = "arXiv",
    primaryClass = "quant-ph",
    doi = "10.1103/myrb-nyhf",
    journal = "PRX Quantum",
    volume = "7",
    number = "1",
    pages = "010344",
    year = "2026"
}

@book{matveev2007algorithmic,
  author          = {Matveev, Sergei},
  title           = {Algorithmic Topology and Classification of 3-Manifolds},
  edition         = {2},
  series          = {Algorithms and Computation in Mathematics},
  volume          = {9},
  publisher       = {Springer},
  address         = {Berlin},
  year            = {2007},
  doi             = {10.1007/978-3-540-45899-9}
}

@article{hou2026stabilizer,
  title={Stabilizer entanglement enhances magic injection},
  author={Hou, Zong-Yue and Cao, ChunJun and Yang, Zhi-Cheng},
  journal={npj Quantum Inf.},
  volume={12},
  pages={113},
  doi={10.1038/s41534-026-01265-4},
  year={2026},
  publisher={Nature Publishing Group UK, London}
}

@article{iannotti2025entanglement,
  title={Entanglement and stabilizer entropies of random bipartite pure quantum states},
  author={Iannotti, Daniele and Esposito, Gianluca and Campos Venuti, Lorenzo and Hamma, Alioscia},
  journal={Quantum},
  volume={9},
  pages={1797},
  year={2025},
  publisher={Verein zur F{\"o}rderung des Open Access Publizierens in den Quantenwissenschaften}
}

@article{jplh-zl35,
  title = {Asymptotically independent fluctuations of stabilizer {R\'enyi} entropy and entanglement in random unitary circuits},
  author = {Szombathy, Dominik and Valli, Angelo and Moca, C{\ifmmode \u{a}\else \u{a}\fi{}}t{\ifmmode \u{a}\else \u{a}\fi{}}lin Pa{\ifmmode \mbox{\c{s}}\else \c{s}\fi{}}cu and Farkas, L{\'o}r{\'a}nt and Zar{\'a}nd, Gergely},
  journal = {Phys. Rev. Res.},
  volume = {7},
  issue = {4},
  pages = {043072},
  numpages = {8},
  year = {2025},
  month = {Oct},
  publisher = {American Physical Society},
  doi = {10.1103/jplh-zl35},
  url = {https://link.aps.org/doi/10.1103/jplh-zl35}
}

@article{PhysRevB.110.045101,
  title = {Nonstabilizerness versus entanglement in matrix product states},
  author = {Frau, M. and Tarabunga, P. S. and Collura, M. and Dalmonte, M. and Tirrito, E.},
  journal = {Phys. Rev. B},
  volume = {110},
  issue = {4},
  pages = {045101},
  numpages = {13},
  year = {2024},
  month = {Jul},
  publisher = {American Physical Society},
  doi = {10.1103/PhysRevB.110.045101},
  url = {https://link.aps.org/doi/10.1103/PhysRevB.110.045101}
}

@article{Haah:2018classification,
    author = "Haah, Jeongwan",
    title = "{Classification of translation invariant topological Pauli stabilizer codes for prime dimensional qudits on two-dimensional lattices}",
    eprint = "1812.11193",
    archivePrefix = "arXiv",
    primaryClass = "quant-ph",
    doi = "10.1063/5.0021068",
    journal = "J. Math. Phys.",
    volume = "62",
    number = "1",
    pages = "012201",
    year = "2021"
}

@misc{Wang:2026decoupling,
    author = "Wang, Yifei and Ni, Zhongyi and He, Mingxin and Liu, Jinguo and Gu, Yingfei",
    title = "{Decoupling 2D translation-invariant topological CSS codes}",
    eprint = "2608.09915",
    archivePrefix = "arXiv",
    year = "2026"
}

@article{Balasubramanian:2024bfb,
    author = "Balasubramanian, Mahesh and Buican, Matthew and Radhakrishnan, Rajath",
    title = "{On the Classification of Bosonic and Fermionic One-Form Symmetries in $2+1$d and 't Hooft Anomaly Matching}",
    eprint = "2408.00866",
    archivePrefix = "arXiv",
    primaryClass = "hep-th",
    doi = "10.1007/s00220-025-05494-0",
    journal = "Commun. Math. Phys.",
    volume = "406",
    pages = "319",
    year = "2025"
}

@article{Kapustin:2019thermal,
    author = "Kapustin, Anton and Spodyneiko, Lev",
    title = "{Thermal Hall conductance and a relative topological invariant of gapped two-dimensional systems}",
    eprint = "1905.06488",
    archivePrefix = "arXiv",
    primaryClass = "cond-mat.str-el",
    doi = "10.1103/PhysRevB.101.045137",
    journal = "Phys. Rev. B",
    volume = "101",
    number = "4",
    pages = "045137",
    year = "2020"
}

@article{Ellison:2021vth,
    author = "Ellison, Tyler D. and Chen, Yu-An and Dua, Arpit and Shirley, Wilbur and Tantivasadakarn, Nathanan and Williamson, Dominic J.",
    title = "{Pauli Stabilizer Models of Twisted Quantum Doubles}",
    eprint = "2112.11394",
    archivePrefix = "arXiv",
    primaryClass = "quant-ph",
    doi = "10.1103/PRXQuantum.3.010353",
    journal = "PRX Quantum",
    volume = "3",
    number = "1",
    pages = "010353",
    year = "2022"
}

\clearpage
\onecolumngrid
\begin{center}
\bfseries End Matter
\end{center}
\vspace{0.6em}

\twocolumngrid

\setcounter{equation}{0}
\renewcommand{\theequation}{A\arabic{equation}}
\renewcommand{\theHequation}{A\arabic{equation}}

\begin{figure}[t]
\centering
\includegraphics[width=.98\linewidth]{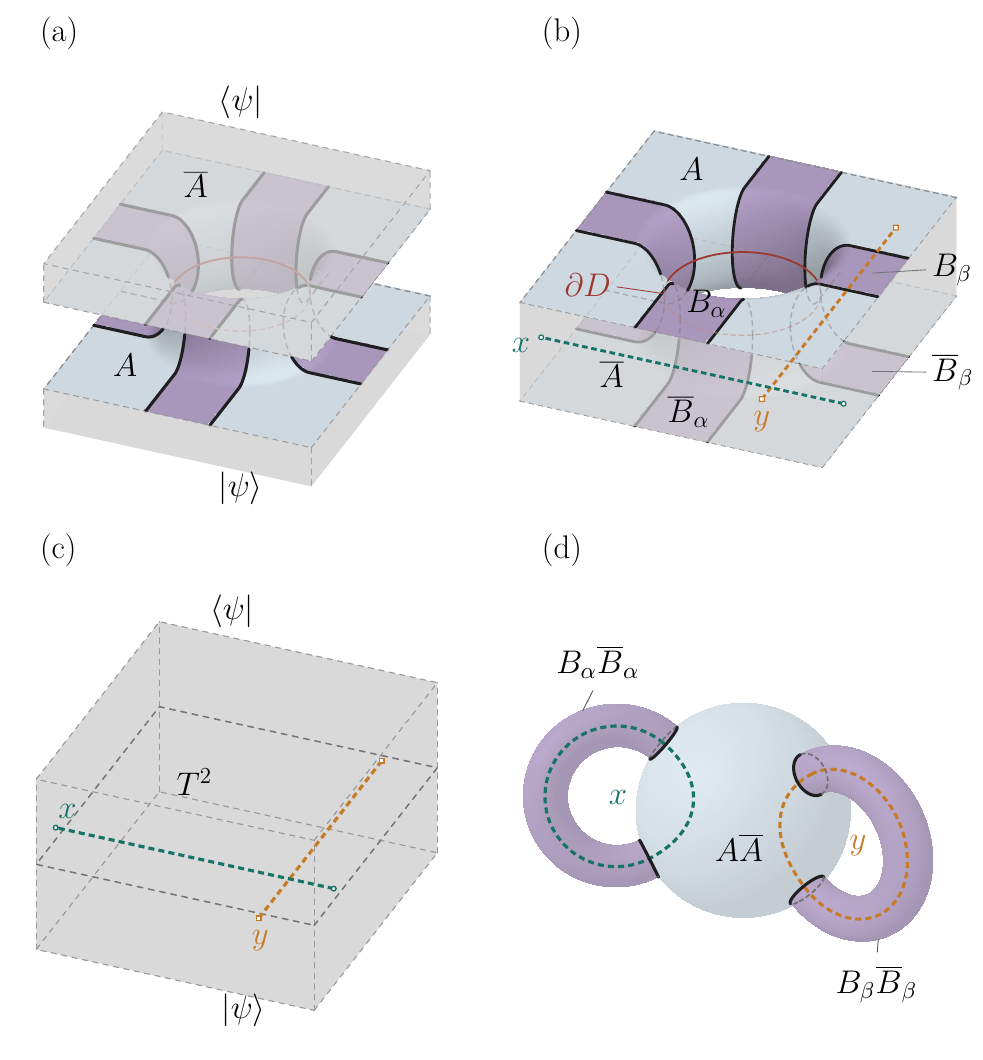}
\caption{Fillings for the Pauli expansion.
(a) Joining a solid-torus filling for $|\psi\rangle$ to its reflection through
the disk bridge $D\times I$ prepares $|\Omega_\psi\rangle$, up to normalization.
The state-preparing insertions are left implicit.
(b) The product filling $\Sigma_{1,1}\times I$ with Wilson lines of charges $x$ and $y$.
(c) Gluing (a) to the reflection of (b) fills the bore and gives the closed
diagram $\langle\psi|P_{xy}|\psi\rangle$.
A dashed outline marks a $T^2$ section; the Wilson lines retain their positions
in the reflected product filling.
(d) The same filling as (b), drawn as a ball with two handles. After normalization,
both views prepare the tube-charge state $|x,y\rangle$.
The boundary regions and Wilson lines use the colors of
Fig.~\ref{fig:skeleton-partition-g1}. Opposite edges of each torus chart are
identified, and matching endpoint markers close the Wilson lines.}
\label{fig:end-fillings}
\end{figure}

\emph{Path-integral derivation of the Pauli expansion.---}
We derive the fact that the amplitudes of $|\Omega_\psi\rangle$ in tube-charge basis $|x,y\rangle$ describe the logical Pauli expectation values, directly from the path-integral in
Fig.~\ref{fig:end-fillings}. We prepare $|\Omega_\psi\rangle$ and $|x,y\rangle$ through path-integral and then compute their overlap by gluing them along the surface.

Choose a solid-torus filling that prepares the normalized ground state
$|\psi\rangle$, leaving its state-preparing insertion implicit.
Joining this filling to its reflected copy along $D$ implements the partial
trace. Vectorization regards the remaining ket and bra surfaces as the
single boundary $\Sigma_2$; the resulting filling thus prepares
$|\Omega_\psi\rangle$ of Eq.~\eqref{eq:toric-vectorization}, up to normalization
[Fig.~\ref{fig:end-fillings}(a)].

The tube-charge basis $\ket{x,y}$ is described by Fig.~\ref{fig:end-fillings}(d), a path-integral on the ball plus two handles that prepares a state on the $\Sigma_2$ surface. The charge $x$ threads one handle and $y$ threads the other. Fig.~\ref{fig:end-fillings}(b) describes the same geometry, represented by the path integral on $\Sigma_{1,1}\times I$.

To compute $\langle x,y|\Omega_\psi\rangle$, glue the filling in
Fig.~\ref{fig:end-fillings}(a) to the reflection of (b) along their common
$\Sigma_2$ boundary, matching the original and reflected copies and their cycles.
The disk bridge fills the
bore of $\Sigma_{1,1}\times I$: in each interval slice, $D$ and
$T^2\setminus D$ restore the full torus. Before normalizing the two preparations,
the resulting closed diagram [Fig.~\ref{fig:end-fillings}(c)] evaluates to
$\langle\psi|P_{xy}|\psi\rangle$, as the Wilson lines inherit the ordering, framing, and overall phase of the Hermitian Pauli operator $P_{xy}$.

To normalize the states, glue each filling to its reflection.
The double of $\Sigma_{1,1}\times I$ is the connected sum of two copies of $S^1\times S^2$.
The boundary Pauli action is unitary, so inserting the Wilson lines does
not change the norm. Using $Z(S^1\times S^2)=1$ and $Z(S^3)=1/2$
for the toric-code TQFT~\cite{dijkgraaf1990gauge}, the connected-sum rule gives
the squared norm of the preparation in Fig.~\ref{fig:end-fillings}(b):
\begin{equation}
\frac{Z(S^1\times S^2)^2}{Z(S^3)}=2.
\label{eq:end-tqft-basis-norm}
\end{equation}
In the double of the filling in Fig.~\ref{fig:end-fillings}(a),
the sewing disk and its reflection form a
sphere that is not crossed by any anyon line. Cutting along this sphere
and capping the two sides leaves two copies of the norm diagram
$\langle\psi|\psi\rangle=1$. Applying the same connected-sum rule gives the squared norm
\begin{equation}
\frac{\langle\psi|\psi\rangle^2}{Z(S^3)}=2.
\label{eq:end-tqft-source-norm}
\end{equation}
Both preparations have norm $\sqrt2$. Dividing the glued amplitude
by the product of these norms yields
\begin{equation}
\langle x,y|\Omega_\psi\rangle
=\frac12\langle\psi|P_{xy}|\psi\rangle.
\label{eq:end-tqft-amplitude}
\end{equation}

\emph{Explicit lattice calculation.---}
Here we confirm $\zeta_n=H_n(p)$, with $p$ given by
Eq.~\eqref{eq:tube-probability}, by an explicit lattice calculation. We first consider the torus. The density matrix separates the local
constraints from the logical information:
\begin{equation}
 |\psi\rangle\langle\psi|=\prod_v\frac{I+A_v}{2}\prod_p\frac{I+B_p}{2}\, L_\psi,
 \label{eq:stabilizer-state}
\end{equation}
where
\begin{equation}
 L_\psi=\frac14\sum_{P\in\calP_2}\ell_P P,
 \quad \ell_P=\langle\psi|P|\psi\rangle,
 \label{eq:stabilizer-projectors}
\end{equation}
with $\calP_2=\{I,X,Y,Z\}^{\otimes 2}$ and $\sum_P|\ell_P|^2=4$ by purity of $\psi$.
Let $\mathcal S$ be the stabilizer group generated by the $A_v$ and $B_p$,
and let $|E|$ be the number of physical qubits (edges). Since
$|\mathcal S|=2^{|E|-2}$, expanding the projectors gives
\begin{equation}
 |\psi\rangle\langle\psi|
 =2^{-|E|}\sum_{s\in\mathcal S}\sum_{P\in\calP_2}\ell_P\,sP.
 \label{eq:end-toric-density}
\end{equation}

Choose string representatives of the logical Pauli operators outside the
disk $D$. Let $\mathcal S_{\Sigma_{1,1}}$ be the subgroup supported in
$\Sigma_{1,1}$, and $|E_{\Sigma_{1,1}}|$ the number of retained qubits. Only
terms acting trivially on $D$ survive the partial trace, and orthogonality
of Pauli operators gives
\begin{equation}
 \begin{aligned}
 \rho&=2^{-|E_{\Sigma_{1,1}}|}\sum_{s\in\mathcal S_{\Sigma_{1,1}}}\sum_{P\in\calP_2}\ell_P\,sP,\\
 \Tr\rho^2&=\frac{4|\mathcal S_{\Sigma_{1,1}}|}{2^{|E_{\Sigma_{1,1}}|}}.
 \end{aligned}
 \label{eq:end-toric-trace}
\end{equation}
This is Eq.~\eqref{eq:rho-pauli} on the lattice, with
$\rho_0=2^{-|E_{\Sigma_{1,1}}|}\sum_{s\in\mathcal S_{\Sigma_{1,1}}}s$, which is proportional
to a projector.
Because $\psi$ is pure, $\rho$ is proportional to a projector
and $\rho/\sqrt{\Tr\rho^2}=\sqrt\rho$. Hence $|\Omega_\psi\rangle$ in
Eq.~\eqref{eq:toric-vectorization} is the canonical purification of $\rho$.

To obtain the Schmidt decomposition of $|\Omega_\psi\rangle$, we group
terms that differ by stabilizers supported entirely in $A$ or in $B$.
Denote these subgroups by $\mathcal S_{A}$ and $\mathcal S_{B}$. There
are $\mathcal N=|\mathcal S_{\Sigma_{1,1}}|/(|\mathcal S_{A}||\mathcal S_{B}|)$
remaining classes. Each class $\sigma$ and Pauli string $P$ give a
normalized vector on each side of the doubled partition, and
\begin{equation}
 |\Omega_\psi\rangle
 =\frac{1}{2\sqrt{\mathcal N}}\sum_{P\in\calP_2}\ell_P
 \sum_{\sigma=1}^{\mathcal N}|P,\sigma\rangle_{A\bar A}|P,\sigma\rangle_{B\bar B}.
 \label{eq:end-toric-schmidt}
\end{equation}
For $P=P_{xy}$, the inner sum divided by $\sqrt{\mathcal N}$ is the tube-charge state
$|x,y\rangle$ of Eq.~\eqref{eq:omega-pauli}, here in its Schmidt form.
These vectors are orthogonal on either side: distinct Pauli restrictions
are orthogonal, and neither $A$ nor the disjoint bands $B$ support a
nontrivial logical operator. Absorbing the signs of $\ell_P$ into the
basis vectors gives the Schmidt form. The nonzero eigenvalues of $\rho_{A\bar A}$
are
\begin{equation}
 \lambda_{P,\sigma}=\frac{p_P}{\mathcal N},\quad \sigma=1,\ldots,\mathcal N,
 \label{eq:end-toric-spectrum}
\end{equation}
with $p_P$ from Eq.~\eqref{eq:pauli-distribution} and sectors of zero weight omitted; for $P=P_{xy}$ this is the weight $p_{xy}$ of Eq.~\eqref{eq:tube-probability}.

At fixed tube charges, the binary boundary degrees of freedom on each
of the four entanglement circles obey one parity constraint, and all
allowed configurations have equal weight. If the $j$th circle has length $L_j$, counted
by the vertices shared by $A\bar A$ and $B\bar B$, it carries $L_j$ such degrees of freedom,
and the total length $L=\sum_jL_j$ is the one in Eq.~\eqref{eq:see-definition}. Then
\begin{equation}
 \mathcal N=\prod_{j=1}^4 2^{L_j-1}=2^{L-4}.
\end{equation}
The boundary spectrum is flat, so $\alpha_n=\log2$ is independent of $n$, and
$\log\mathcal N=\alpha_n L-4\log2$. Consequently,
\begin{equation}
 S_n(\rho_{A\bar A})=\alpha_n L-4\log2+H_n(p).
 \label{eq:end-toric-entropy}
\end{equation}
Comparing with Eq.~\eqref{eq:see-definition}, with $\gamma=\log2$, gives
$\zeta_n=H_n(p)$, in agreement with the topological argument, including
the von Neumann limit $n\to1$.

For a genus-$g$ surface, the same calculation applies to $2g$ logical
qubits. We remove the single disk $D$, replace $\Sigma_{1,1}$ by
$\Sigma_{g,1}$, and use $L_\psi=4^{-g}\sum_{P\in\calP_{2g}}\ell_P P$. The partial trace retains the
stabilizers supported in $\Sigma_{g,1}$, with
$\Tr\rho^2=4^g|\mathcal S_{\Sigma_{g,1}}|/2^{|E_{\Sigma_{g,1}}|}$, and
$\rho$ is again proportional to a projector. The disk $A$ and the $2g$
disjoint bands support no nontrivial logical operators, so the same
grouping applies, with the prefactor $1/(2\sqrt{\mathcal N})$ in
Eq.~\eqref{eq:end-toric-schmidt} replaced by $1/(2^g\sqrt{\mathcal N_g})$.
With $4g$ entanglement circles,
$\mathcal N_g=\prod_{j=1}^{4g}2^{L_j-1}=2^{L-4g}$, so
$\log\mathcal N_g=\alpha_n L-4g\log2$. The eigenvalues are
$p_P/\mathcal N_g$, each repeated $\mathcal N_g$ times, with
$p_P=4^{-g}|\ell_P|^2$, and
\begin{equation}
 S_n(\rho_{A\bar A})=\alpha_n L-4g\log2+H_n(p).
 \label{eq:end-toric-genus}
\end{equation}
This reproduces Eq.~\eqref{eq:toric-higher-genus}.

\end{document}